\documentclass[fleqn,twoside]{article}
\usepackage{amssymb}
\usepackage{amsbsy}
\usepackage{amsfonts}
\usepackage{graphics}
\usepackage{epsfig}
\usepackage{latexsym}
\usepackage[dvips]{color}
\usepackage{espcrc2}
\title{\vspace{-2.0cm}
       {\normalsize DESY 02--147}    \\[-0.2cm]
       {\normalsize Edinburgh 2002/09}   \\[-0.2cm]
       {\normalsize LU-ITP 2002/018}   \\[0.35cm]
        Structure functions near the chiral limit\thanks{Poster at Lattice 
2002, Cambridge, USA.}\\[0.3em]}

\author{M. G\"ockeler\address{Institut f\"ur Theoretische Physik, 
Universit\"at Leipzig, D-04109 Leipzig, Germany}$^,$\address{Institut f\"ur 
Theoretische Physik, Universit\"at Regensburg, D-93040 Regensburg, Germany},
R. Horsley\address{School of Physics, University of
Edinburgh, Edinburgh EH9 3JZ, UK}, D. Pleiter\address{Deutsches 
Elektronen-Synchrotron DESY, John von Neumann-Institut f\"ur Computing NIC,\\
\hspace{0.173cm}D-15735 Zeuthen, Germany}, P.E.L Rakow\address{Theoretical
Physics Division,  
Department of Mathematical Sciences, University of Liverpool,\\
\hspace{0.173cm}Liverpool  
L69 3BX, UK}, and G. Schierholz$^{\rm d,}$\address{Deutsches 
Elektronen-Synchrotron DESY, D-22603 Hamburg, Germany}}

\begin{document}

\begin{abstract}
We compute hadron masses and the lowest moments of unpolarized and 
polarized nucleon structure functions down to pion masses of $300 \, 
\mbox{MeV}$, in an effort to make unambiguous predictions at the physical 
light quark mass.
\end{abstract}

\maketitle

\section{INTRODUCTION} 

Understanding the structure of hadrons in terms of quark and gluon 
constituents (partons), in particular how quarks and gluons provide the 
binding and spin of the nucleon, is one of the outstanding problems in 
particle physics.

Moments of parton distribution functions, such as $\langle x \rangle$ and $g_A
= \Delta\,u -\Delta\,d$, are benchmark calculations in lattice QCD. To compare
the lattice results with experiment, one must extrapolate the data from the
lowest calculated quark mass to the physical value. Up to now, most results
are for quark masses of about the strange quark mass and larger. A naive,
linear extrapolation of $\langle x \rangle$ in the quark mass overestimates 
the experimental
number by $\approx$ 40\,\%~\cite{QCDSF1}. While most results are for quenched 
QCD, where
one might expect that $\langle x \rangle$ is larger than the experimental
value, recent unquenched results~\cite{QCDSF2,Sesam,GS} indicate that this
problem remains. 

It has been argued that a linear extrapolation must fail because it omits
non-analytic structure associated with chiral symmetry breaking~\cite{chiral}.
Indeed, chiral perturbation theory~\cite{chiral,chiral2}
suggests a large deviation of $\langle x \rangle$ and $g_A$ from linearity as
the pion (quark) mass tends to zero:
\begin{eqnarray}
\langle x \rangle_{NS} &\!\!\!\!\!=\!\!\!\!& \langle x \rangle_{NS}^0 
\Big(1-\frac{3 g_A^2+1}
{(4\pi f_\pi)^2} m_\pi^2 \ln\big(\frac{m_\pi^2}{m_\pi^2 +\Lambda^2}\big)\Big)
\nonumber \\
& & \hspace*{3.5cm}+\, O(m_\pi^2),\\
g_A &\!\!\!\!\!=\!\!\!\!& g_A^0 \Big(1-\frac{2 g_A^2+1}
{(4\pi f_\pi)^2} m_\pi^2 \ln\big(\frac{m_\pi^2}{m_\pi^2+\Lambda^2}\big)\Big)
\nonumber\\
& & \hspace*{3.5cm}+\, O(m_\pi^2),
\end{eqnarray}
where $\Lambda$ is a phenomenological parameter. Equation (1) fits both the 
lattice data and the 
experimental value~\cite{chiral}. However, $\Lambda$ is not yet determined by 
the lattice data. To constrain this parameter, and to perform an accurate 
extrapolation based solely on lattice results, data at smaller quark masses 
are crucial.

\section{THE SIMULATION}

To narrow the gap between the `chiral' regime and the lowest 
calculated quark mass, we have started simulations at quark masses 
corresponding to about twice the physical pion mass. The calculations are 
done for Wilson fermions in the quenched approximation. Improved Wilson 
fermions are known to suffer from exceptional configurations, which would 
forbid such a calculation at currently accessible couplings.

We work at $\beta = 6.0$. Our present data sample consists of:
\vspace*{0.25cm}

\begin{center}
\begin{tabular}{c|c|c|c} 
$\beta$ & $\kappa$ & Volume & Configurations \\ \hline
6.0 & 0.1515 & $16^3\,32$ & O(5000) \\
6.0 & 0.1530 & $16^3\,32$ & O(5000) \\
6.0 & 0.1550 & $16^3\,32$ & O(5000) \\ \hline
6.0 & 0.1550 & $24^3\,32$ & 220 \\ 
6.0 & 0.1558 & $24^3\,32$ & 220 \\ 
6.0 & 0.1563 & $24^3\,32$ & 220 \\ \hline
6.0 & 0.1563 & $32^3\,48$ & 50 \\
6.0 & 0.1566 & $32^3\,48$ & 100 \\ 
\end{tabular}
\end{center}
\vspace*{0.25cm}

\noindent
The $24^3\,32$ and $16^3\,32$ lattices are from our previous 
runs~\cite{scaling}. On all our lattices $m_\pi\,L > 4$, so that finite volume 
effects may be expected to be small. We did not see any at $\kappa = 0.1563$.
We are currently increasing our statistics at the smaller quark masses.

\section{HADRON MASSES}

We first looked at the chiral behavior of pion, $\rho$ and nucleon masses. We 
did not find compelling evidence for non-analytic behavior in any of the three
cases. In Fig.~1 we plot the nucleon mass as a function of the pion mass. 
Quenched chiral perturbation theory (qCPT) predicts~\cite{qcpt}
\begin{equation}
m_N = m_N^0 + C_{1/2} m_\pi + C_1 m_\pi^2 + C_{3/2} m_\pi^3
\end{equation}
with 
\begin{equation}
C_{1/2}= - \frac{3\pi}{2} \big(D-3F\big)^2\delta \approx - 0.5.
\end{equation}
Empirically one finds, to a very high precision,
\begin{equation}
m_N = \sqrt{m_N^{0\;2} + C_1 m_\pi^2},
\end{equation}
which is an analytic function in the quark mass. In Fig.~1 we fit (3) and (5)
to the data. Both fits are hardly distinguishable in the range of the data. 
Fit (3) gives $C_{1/2} = 0.7(1)$ and 
$\chi^2 = 0.8$. Note that the coefficient $C_{1/2}$ comes out 
positive instead of negative as it should, according to qCPT. There is some 
preference for the single parameter analytic fit (5), which gives
$\chi^2 = 0.4$. What comes as a surprise is that the resulting physical 
nucleon masses (denoted by open circles in Fig.~1) still differ by more than 
10 \%. 
We conclude that pion masses of 300 MeV are not small 
enough to sufficiently constrain the fit function, so as to give unambiguous
results.

\begin{figure}[tbp]
\vspace*{-0.1cm}
  \begin{center}
    \epsfig{file=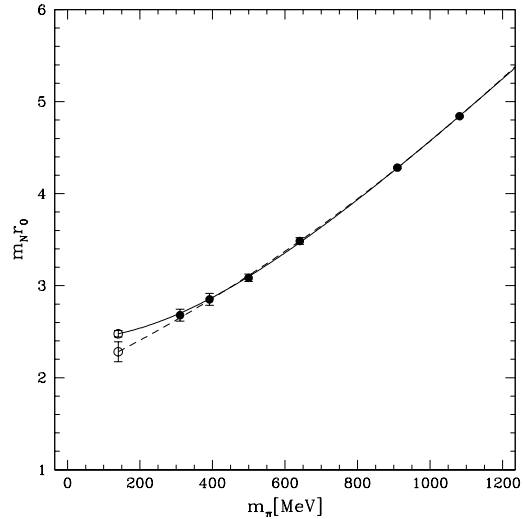,height=7.3cm,width=7.3cm}
  \end{center}
\vspace*{-1.20cm}
\caption{The nucleon mass against $m_\pi$, together with the phenomenological
fit (5) (solid curve) and the chiral fit (3) (dashed curve). We have taken the 
nucleon mass at the physical pion mass to set the scale, using (3).}
\vspace*{-0.55cm}
\end{figure}

\section{STRUCTURE FUNCTIONS}

Let us now turn to the moments of non-singlet, unpolarized and polarized
structure functions. We consider $\langle x \rangle_{NS}$ first. On the 
lattice one does not compute $\langle x \rangle_{NS}$ directly 
but~\cite{QCDSF1}
\begin{equation}
R = \langle x \rangle_{NS}\, m_N,
\end{equation}
which can be interpreted as the fraction of the proton's mass carried by the 
$u$ quark minus that carried by the $d$ quark. Because chiral perturbation 
theory does not provide much guidance at present quark masses, we seek a 
phenomenological extrapolation.
In Fig.~2 we plot $R$ as a function of the pion mass. The data points have 
been properly renormalized and converted to renormalization group invariant 
(RGI) numbers following~\cite{QCDSF2}.
We find that the data lie precisely on a straight line, all the way from 
$m_\pi \approx 1 \, \mbox{GeV}$ down to our smallest pion mass at $m_\pi = 300
\, \mbox{MeV}$. The slope comes out to be $\approx 0.4$, which is close to the 
value $1/3$ expected in the heavy quark limit. A linear extrapolation
to the physical pion mass, $R = R^0 + R_{1/2}\, m_\pi$, gives a value for
$\langle x\rangle_{NS}^{RGI}$ that is much closer to the experimental 
number than previous results.
A fit of (1) to $\langle x \rangle_{NS}$ still does not constrain $\Lambda$. We
find $\Lambda = 360^{+120}_{-360} \; \mbox{MeV}$. 
Our present data cannot distinguish between the fit function (1) 
and $(R^0 + R_{1/2}\, m_\pi)/\sqrt{m_N^{0\;2} + C_1 m_\pi^2}$.

\begin{figure}[tbp]
\vspace*{-0.1cm}
  \begin{center}
    \epsfig{file=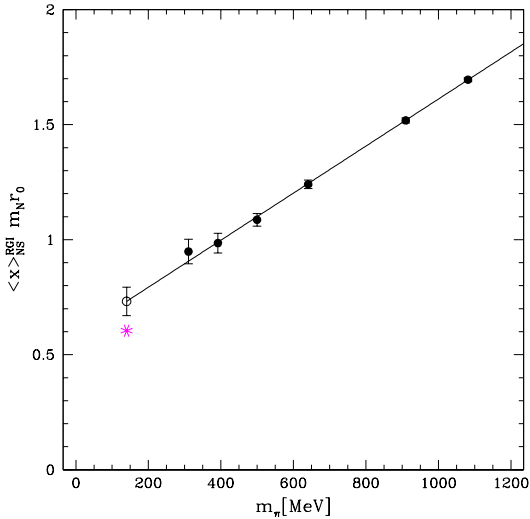,height=7.3cm,width=7.3cm}
  \end{center}
\vspace*{-1.20cm}
\caption{The lowest moment of the non-singlet, unpolarized structure function 
against $m_\pi$, together with a linear extrapolation to the physical pion 
mass. Also shown is the experimental result ($+\hspace{-7.65pt}\times$) taken 
from~\cite{MRS}.}
\vspace*{-0.55cm}
\end{figure}

Let us now consider $g_A$, the axial vector coupling of the nucleon. This
quantity describes the fraction of the proton's spin 
carried by the $u$ quark minus that carried by the $d$ quark. In Fig.~3 we
plot $g_A$ as a function of the pion mass. Again, the data lie on a straight
line, and we do not see any sign of non-analytic behavior as suggested by (2).

\begin{figure}[tbp]
\vspace*{-0.11cm}
  \begin{center}
    \epsfig{file=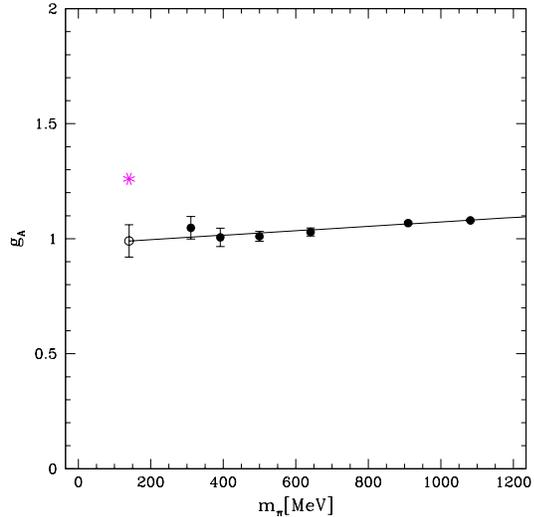,height=7.3cm,width=7.3cm}
  \end{center}
\vspace*{-1.20cm}
\caption{The axial vector coupling of the nucleon against $m_\pi$, together
with a linear extrapolation and the experimental number 
($+\hspace{-7.65pt}\times$).}
\vspace*{-0.55cm} 
\end{figure}

\section{CONCLUSION}

It appears that at pion masses of 300 MeV we are not 
yet sensitive to the predictions of chiral perturbation theory. Whether the 
situation will improve on finer lattices has to be seen.
\vspace*{0.25cm}

\noindent
The numerical calculations were performed on the APE100 at NIC (Zeuthen) and 
on the Cray T3E at NIC (J\"ulich).


\begin{thebibliography}{9}
\bibitem{QCDSF1}
M. G\"ockeler et al., Phys. Rev. D53 (1996) 2317.
\bibitem{QCDSF2} S. Capitani et al., Nucl. Phys. B (Proc. Suppl.) 106 
(2002) 299.
\bibitem{Sesam}
D. Dolgov et al., {\tt hep-lat/0201021}.
\bibitem{GS}
R. Horsley, this conference.
\bibitem{chiral}
W. Detmold et al., Phys. Rev. Lett. 87 (2001) 172001.
\bibitem{chiral2}
J.-W. Chen and X. Ji, Phys. Rev. Lett. 88 (2002) 052003;
D. Arndt and M.J. Savage, Nucl. Phys. A697 (2002) 429.
\bibitem{scaling}
M. G\"ockeler et al., Phys. Rev. D57 (1998) 5562.
\bibitem{qcpt}
C. Bernard and M.F.L. Golterman, Phys. Rev. D46 (1994) 853;
J.N. Labrenz and S.R. Sharpe, Phys. Rev. D54 (1996) 4595. 
\bibitem{MRS}
A. Martin, R.G. Roberts and W.J. Stirling, Phys. Lett. B354 (1995) 155.
\end{thebibliography}
\end{document}